# *USRN Discovery Pilot: Increasing the Discoverability of Open Access Content Through a National Network*

*Petr Knoth, CORE, KMi, The Open University,* petr.knoth@open.ac.uk *ORCID:* 0000-0003-1161-7359

*Paul Walk, Antleaf,* paul@paulwalk.net

*Matteo Cancellieri, CORE, KMi, The Open University,* matteo.cancellieri@open.ac.uk *ORCID:* 0000-0002-9558-9772

*Michael Upshall, CORE, KMi, The Open University,* michael.upshall@open.ac.uk

*Halyna Torchylo, CORE, KMi, The Open University,* halyna.torchylo@open.ac.uk

*Jennifer Beamer, SPARC,* jennifer@sparcopen.org

*Kathleen Shearer, COAR,* m.kathleen.shearer@gmail.com

*Heather Joseph, SPARC,* heather@sparcopen.org

## Abstract

This paper presents the results of the USRN Discovery Pilot Project, a collaboration of SPARC, the Confederation of Open Access Repositories (COAR), CORE and Antleaf, to enhance the discoverability of research papers in US repositories leveraging CORE as an indexing service for USRN repositories.

The project conducted actions in three strategic areas:

- Assessing and quantitatively measuring discoverability and barriers to it at the beginning and end of the pilot project,
- conducting interventions to increase discoverability, and
- supporting interventions by technology and guidelines (provided by CORE services), to minimise effort and maximise effect.

The key results of the project include:

- Around three-quarters of a million research outputs held in the selected US repositories have been made discoverable (a 50% increase) compared to the year before;
- The project has made available the CORE Data Provider's Guide as well as a selection of new and improved tools to support repositories in increasing their discoverability. These include the CORE Reindexing Button and Index Notification modules, Fresh Finds and the USRN Desirable Characteristics for Digital Publication Repositories checking tool.

The project team is now exploring ways to scale out this work to include more repositories.

# Introduction

The USRN Discovery Pilot Project was initiated to address a fundamental challenge in the scholarly communication ecosystem: ensuring that content deposited in repositories is discoverable. Repositories are a cornerstone of open-access infrastructure, yet many in the US face significant limitations. Unlike counterparts in regions such as the UK, US repositories often lack the resources and technical capacity to ensure comprehensive capture and discoverability of their institution's research outputs. This pilot highlighted these structural challenges, while also showcasing actionable strategies to overcome them.

The project ran for twelve months, from November 2023 to October 2024. Twenty-three institutional repositories were invited to the project based on their contrasting size, activity, and location; and decided to take part in the pilot project. Each of these repositories was then assessed by means of automated tests and using an interview method to reveal and quantify limitations for their discoverability. Each repository was also made aware of the CORE Data Provider's Guide[2] and given access to its CORE Dashboard, which contains a range of tools to help the repository identify and track potential technical issues.

At the start of the project, 15 of the 23 participating repositories had non-functional or partially functional OAI-PMH endpoints, preventing efficient metadata harvesting. Through technical interventions, by the end of the pilot, all but one had operational endpoints. This and other improvements led to a remarkable increase in indexed research outputs, **growing from 1.45 million to over 2.18 million**—a ~50% increase in discoverable content. Table 1 in the Appendix includes a detailed breakdown of these improvements on a per-repository basis.

## Methodology

### Enhancing Repository Management with CORE Tools and USRN Standards

The CORE Dashboard [3] was used as part of the project as an essential interface containing a range of new or improved tools for repository managers, helping them to check and increase the discoverability of content in their repository. These included:

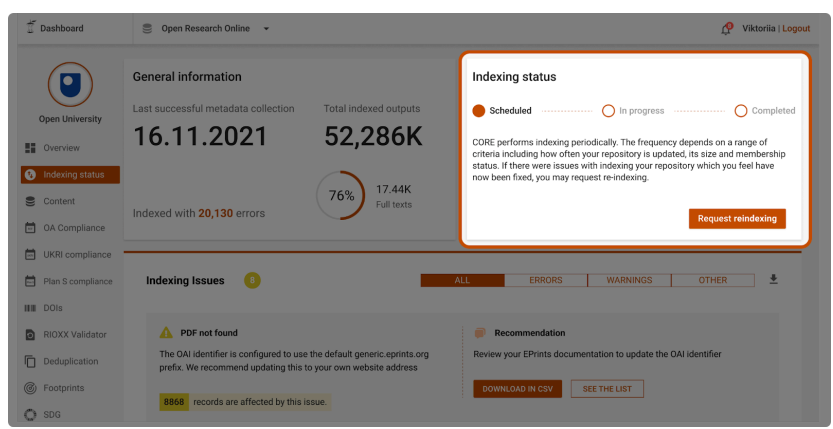

**Reindexing Tools**: A "request harvesting" button allows institutions to request reindexing of their research outputs. Additionally, an email notification service informs repository managers of reindexing events, helping to monitor repository visibility and ensure smooth OAI-PMH operations.

Figure 1: The request reharvesting feature embedded in the CORE Dashboard.

**Rights Retention Statement Checker**: This tool uses machine learning to locate and display rights-related text (e.g., "CC-BY") in articles, reducing manual efforts by ~75% and improving rights management efficiency.

**Fresh Finds Prototype**: A new pilot module designed to track newly published articles by institution-affiliated academics, this tool highlights works missing from the repository, lowering barriers to maintaining comprehensive records of institutional outputs.

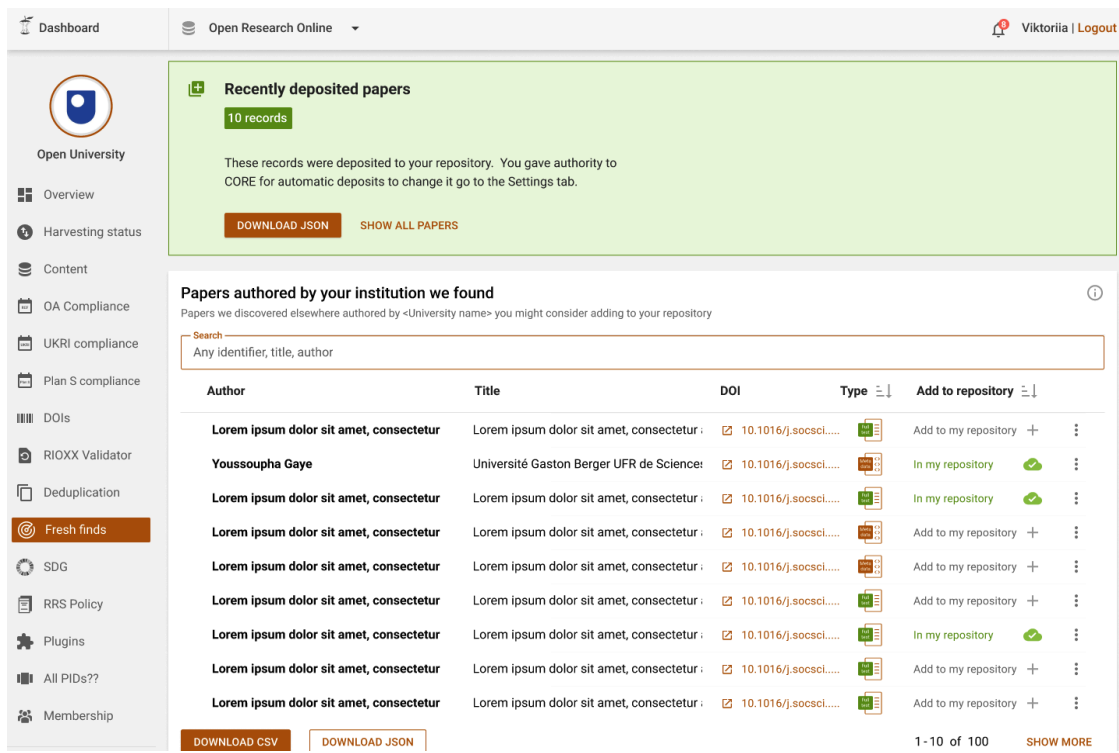

Figure 2: Fresh Finds prototype

**USRN Desirable Characteristics Module**: This module assesses repository adherence to the USRN Desirable Characteristics [1] —best practices for repositories, including persistent identifiers, high-quality metadata, long-term sustainability, and provenance tracking. It provides actionable recommendations with compliance scores for each characteristic.

Figure 3: A screenshot of the USRN Desirable Characteristics automated checking tool

**USRN Desirable Characteristics Toolkit [4]:** A wiki-based resource complements the Dashboard by detailing adoption strategies, linking to related tools, and offering resources aligned with specific desirable characteristics like free access, metadata quality, and common formats. Hosted on GitHub, it encourages community contributions for ongoing refinement. Together, these innovations promote sustainable, discoverable, and standards-compliant repositories.

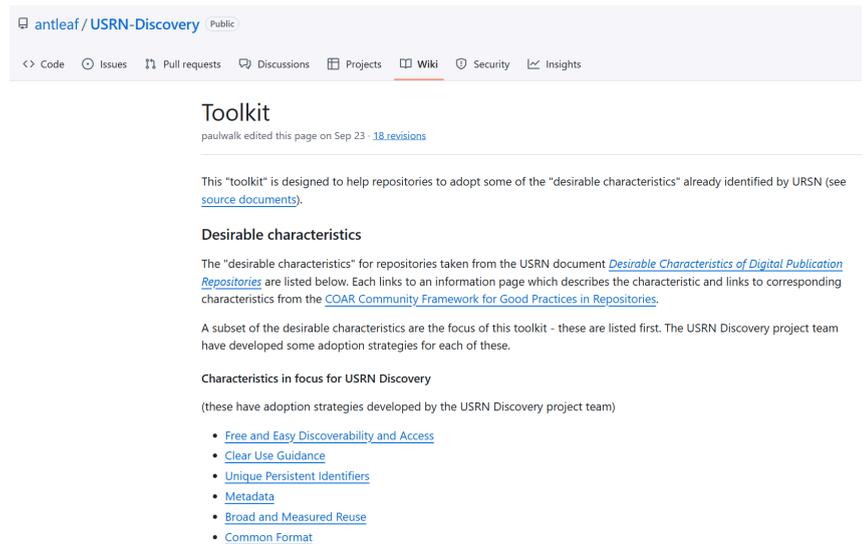

*Figure 4: The USRN Desirable Characteristics Toolkit designed to provide advice to repositories on how to comply with best practices*

## Discussion

During this project, we identified an unexpectedly large number of repositories that have either never configured their OAI-PMH harvesting correctly, or that have over time changed their platform software including their OAI-PMH configuration, which inadvertently resulted in indexing services no longer being able to access their content.

Given the lack of public awareness of repositories, there would be a significant benefit in presenting regular metrics that indicate the continuing health of repositories to a wider audience. This could be done by tracking (for example) continuing growth in the number of objects held by repositories, or by other non-attributable indicators that provided indexed data.  CORE can provide these metrics, anonymised to show growth using demographic data only (e.g. US repositories increased in size by x% during 2023).

All 23 project partners reported a lack of time to manage their repositories effectively. Large or small, the repositories experienced this issue: they have insufficient time to correct the entries they curate, let alone look for new content to add. This responsibility is frequently left to academics uploading their content, which results in inefficiencies since many academics have low awareness of the importance of metadata standards, and content requires further curation before being made discoverable. CORE's solution to this is to identify efficiencies, largely by building automated tools to identify missing metadata and then fill in the gaps from elsewhere. If this can be effectively provided, then CORE (and the repository networks) will be enabling repository staff to work more efficiently, while at the same time improving metadata quality.

## Conclusion

Beyond technical improvements, the project fostered a deeper understanding of repository management dynamics through surveys and direct engagement with participating institutions. It became evident that many repositories in the US view their primary role as increasing discoverability rather than actively tracking institutional research outputs. This contrasts with practices in regions where institutional mandates drive comprehensive research tracking. Furthermore, we identified a significant awareness gap around essential technical standards and protocols, underscoring the need for targeted education and advocacy.

The pilot also explored the readiness of repositories to comply with forthcoming US open-access policies, particularly the OSTP Nelson Memo. The findings advocated for institutional repositories to be recognised as "designated repositories" based on technical compliance rather than arbitrary designations. This perspective challenges existing assumptions and emphasises the potential of well-maintained institutional repositories to meet federal policy requirements effectively.

## References.

[1] USRN Desirable Characteristics of Digital Publication Repositories - https://sparcopen.org/wp-content/uploads/2022/10/Desirable-Characteristics-of-Digital-Publication-Repositories-APPROVED-20230331.pdf

[2] CORE Data Providers Guidelines https://core.ac.uk/documentation/data-providers-guide

[3] CORE Repository Dashboard https://core.ac.uk/services/repository-dashboard

[4] USRN Discovery toolkit https://github.com/antleaf/USRN-Discovery/wiki/Toolkit

# Appendix

| Institution | Public or private | Carnegie Ranking | Repo software | OAI-PMH endpoint status at project start | Status at project end | Content exposed by CORE at project start | Content exposed by CORE at project end |
|---|---|---|---|---|---|---|---|
| AgEcon (Univ of Minnesota) | n/a | n/a | TIND | Non-operating OAI-PMH | Functional | 0 | 194,000 |
| Berkeley Law Library | public | R1 | TIND | Non-operating OAI-PMH | Functional | 0 | 19,295 |
| Case Western University | private | R1 | BePress | Functional | Functional | 12,664 | 13,260 |
| Columbia University | private` | R1 | Blacklight, Fedora | Functional | Functional | 39,220 | 44,000 |
| CSUSB (California State University, San Bernardino) | Public | R2 | Digital Commons | Functional | Functional | 18,997 | 19,734 |

| Institution | Type | Classification | Platform | OAI-PMH Status | Functional Status | Col 7 | Col 8 |
|---|---|---|---|---|---|---|---|
| Harvard University | private | R1 | DSpace | Non-operating OAI-PMH | Functional | 0 | 46,220 |
| MediArXiv | n/a | n/a | OSF | Did not have OAI-PMH | Functional | 0 | 248 |
| Oregon State University | public | R1 | Samvera Hyrax | Non-operating OAI-PMH | Functional | 0 | 33,932 |
| Princeton University | private | R1 | unknown | Did not have OAI-PMH | Functional | 0 | 9,406 |
| Stanford University | private | R1 | unknown | No OAI-PMH | No OAI-PMH | 0 | 0 |
| Texas A&M University (Texas Digital LIbrary) | public | R1 | DSpace | Non-operating OAI-PMH | Functional | 0 | 118,632 |
| University of Arizona | Public | R1 | other | Little full-text indexing | Functional | 103,740 | 108,205 |
| University of California (eScholarship) | Public | R1 | other | No full-text harvesting | Functional | 420,000 | 440,000 |
| University of Chicago | Private | R1 | Samvera Hyrax | Wrong OAI resolver | Functional | 8371 | 11,360 |
| University of Colorado Boulder | public | R1 | Samvera Hyrax | Non-operating OAI-PMH | Functionarl | 6,220 | 19,119 |
| University of Illinois Chicago | public | R1 | Figshare | Functioning | Functioning | 3600 | 4,760 |

| Institution | Type | Classification | Platform | Initial Status | Final Status | Initial Count | Final Count |
|---|---|---|---|---|---|---|---|
| University of Massachusetts Amherst | Public | R1 | Digital Commons (now DSpace) | Non-operating OAI-PMH | Functioning | 74,740 | 73,301 |
| Miami University, OH | public | R2 | DSpace | Non-operating OAI-PMH | Functioning | 0 | 141,372 |
| University of Michigan | public | R1 | bespoke | Functioning | Functioning | 166,135 | 14,064 |
| University of North Texas - Denton | public | R1 | unknown | Functioning | Functioning | 497,670 | 722,000 |
| University of Texas, Austin | public | R1 | DSpace 7 | Functioning | Functioning | 105,477 | 118,632 |
| Washington State University Worcester Polytechnic Institute (WPI) | private | R2 | ArchiveSpace, 4 Samvera | no full-text harvesting | Functioning | 0 | 34,064 |
| **Totals** | | | | | | **1,456,834** | **2,185,604** |

*Table 1: USRN repository participants and their status at the start and the end of the pilot project.*